\documentclass[conference]{IEEEtran}
\IEEEoverridecommandlockouts

\usepackage{amsmath,amssymb,amsfonts}
\usepackage{array}
\usepackage{graphicx}
\usepackage{latexsym}
\usepackage{setspace}
\usepackage{mathtools}
\usepackage{amsmath}
\usepackage{yhmath}
\usepackage{graphicx,dblfloatfix}
\usepackage{cite}
\usepackage[caption=false,labelformat=simple,font=footnotesize,position=bottom]{subfig}
\usepackage{setspace}
\usepackage{bm}
\usepackage{upgreek}
\usepackage{multirow}
\usepackage{afterpage}
\usepackage{makecell}
\usepackage[T1]{fontenc}
\usepackage{booktabs}
\usepackage{threeparttable}
\usepackage[table]{xcolor}
\usepackage{arydshln}
\usepackage{enumitem}
\usepackage{bbding}
\usepackage{amssymb}
\usepackage{xurl}
\usepackage[breaklinks=true,hidelinks]{hyperref}
\usepackage[figure,figure*]{hypcap}

\DeclareRobustCommand{\approachdef}[1]{\raisebox{\baselineskip}[0pt][0pt]{\hypertarget{def:#1}{\strut}}#1}
\DeclareRobustCommand{\approachref}[1]{\hyperlink{def:#1}{#1}}
\DeclareRobustCommand{\approachreftext}[2]{\hyperlink{def:#1}{#2}}
\DeclareRobustCommand{\acronymdef}[2]{\raisebox{\baselineskip}[0pt][0pt]{\hypertarget{acro:#1}{\strut}}#2}
\DeclareRobustCommand{\acronymref}[2]{\hyperlink{acro:#1}{#2}}
\DeclareRobustCommand{\acronymabsdef}[2]{\raisebox{\baselineskip}[0pt][0pt]{\hypertarget{acroabs:#1}{\strut}}#2}
\DeclareRobustCommand{\acronymabsref}[2]{\hyperlink{acroabs:#1}{#2}}
\usepackage{amsbsy}
\usepackage{tikz}
\usepackage{eso-pic} 
\def\BibTeX{{\rm B\kern-.05em{\sc i\kern-.025em b}\kern-.08em
    T\kern-.1667em\lower.7ex\hbox{E}\kern-.125emX}}

\usetikzlibrary{arrows.meta, positioning, shapes.geometric}

\usepackage{algpseudocode}

\definecolor{myred}{rgb}{0.7686,0.0588,0.0941}
\definecolor{mygreen}{rgb}{0.2000,0.5804,0.1804}
\definecolor{myyellow}{rgb}{0.9,0.8,0}
\definecolor{myyellow2}{rgb}{0.8,0.6,0}
\definecolor{mygray}{rgb}{0.92,0.92,0.92}
\definecolor{mygray2}{rgb}{0.8,0.8,0.8}
\definecolor{mymagenta}{rgb}{0.9999,0.5,0.9999}
\definecolor{mymagenta2}{rgb}{0.7,0.2,0.7}
\definecolor{myGreen}{rgb}{0,0.9,0.7}
\definecolor{myGreen2}{rgb}{0,0.65,0.45}
\definecolor{mygreen3}{rgb}{0.76,0.83,0.61}
\definecolor{myblue}{rgb}{0.6,0.6,1}
\definecolor{myblue2}{rgb}{0.3,0.3,0.9}
\definecolor{myblue3}{rgb}{0.73,0.8,0.90}
\definecolor{myorange}{rgb}{0.9804,0.7137,0.4941}
\definecolor{myorange2}{rgb}{0.9922,0.9020,0.8275}
\definecolor{myorange3}{rgb}{0.99, 0.84,0.71}
\definecolor{bluebl}{HTML}{DBDBDB}

 \newcommand{\Unit}[1]{\ensuremath{{\:}\mathrm{#1}}}

\newcommand{\A}{\Unit{A}}

\newcommand{\V}{\Unit{V}}

\newcommand{\mH}{\Unit{mH}}

\newcommand{\ohm}{\Unit{{\Omega}}}

\newcommand{\Nm}{\Unit{Nm}}
\newcommand{\rpm}{\Unit{r/min}}

\newcommand{\MB}{\Unit{MB}}
\newcommand{\kB}{\Unit{kB}}
\newcommand{\B}{\Unit{B}}

\makeatletter
\newif\ifnobrackets
\renewcommand\@cite[2]{\ifnobrackets\else[\fi{#1\if@tempswa , #2\fi}\ifnobrackets\else]\fi\nobracketsfalse}

\makeatother

\makeatletter
\newcommand{\raisemath}[1]{\mathpalette{\raisem@th{#1}}}
\newcommand{\raisem@th}[3]{\raisebox{#1}{$#2#3$}}
\makeatother

\makeatletter
\renewcommand{\Function}[2]{%
  \csname ALG@cmd@\ALG@L @Function\endcsname{#1}{#2}%
  \def\jayden@currentfunction{#1}%
}
\newcommand{\funclabel}[1]{%
  \@bsphack
  \protected@write\@auxout{}{%
    \string\newlabel{#1}{{\jayden@currentfunction}{\thepage}}%
  }%
  \@esphack
}
\makeatother

\graphicspath{{Figs/}}

\makeatletter
\def\blfootnote{\xdef\@thefnmark{}\@footnotetext}
\makeatother

\makeatletter
\def\ignore#1{%
     \begingroup
         \@fileswfalse
         #1
    \endgroup
}
\makeatother

\makeatletter
\newcommand{\linebreakand}{%
  \end{@IEEEauthorhalign}
  \hfill\mbox{}\par
  \mbox{}\hfill\begin{@IEEEauthorhalign}
}
\makeatother

\begin{document}

\title{Optimization of Current Lookup Tables for Minimum Stator Copper Loss and Torque Ripple in the Full Torque-Speed Range for a Six-Phase PMSM With Nonsinusoidal Back-EMF\\
\thanks{This work was supported in part by the Spanish State Research Agency (AEI) under project PID2022-136908OBI00/MCIN/AEI/10.13039/501100011033/FEDER-UE.}
}

\author{
\IEEEauthorblockN{Alejandro G. Yepes\textsuperscript{1}, Shirin Rahmanpour\textsuperscript{2}, Oscar López\textsuperscript{1}, Wessam E. Abdel-Azim\textsuperscript{1,3}, \\ Petros Karamanakos\textsuperscript{2}, Ayman S. Abdel-Khalik\textsuperscript{4}, and Jesús Doval-Gandoy\textsuperscript{1}}\vspace{3pt}

\IEEEauthorblockA{\textsuperscript{1}\textit{CINTECX}, \textit{Universidade de Vigo}, \textit{APET}, Vigo, Spain}

\IEEEauthorblockA{\textsuperscript{2}\textit{Faculty of Information Technology and Communication Sciences}, \textit{Tampere University}, Tampere, Finland} 

\IEEEauthorblockA{\textsuperscript{3}\textit{Department of Electrical Engineering}, \textit{Alexandria University}, Alexandria, Egypt}

\IEEEauthorblockA{\textsuperscript{4}\textit{Department of Electrical and Computer Engineering}, \textit{Sultan Qaboos University}, Muscat, Oman}

\IEEEauthorblockA{Email: agyepes@uvigo.es, shirin.rahmanpour@tuni.fi, olopez@uvigo.es, wessam.essam@uvigo.es, \\ p.karamanakos@ieee.org, a.abdelkhalik@squ.edu.om, jdoval@uvigo.es}\vspace{-11pt}
}

\maketitle

\AddToShipoutPictureFG*{%
  \AtPageUpperLeft{%
    \raisebox{-5mm}[0pt][0pt]{%
      \makebox[\paperwidth][c]{%
        \scriptsize\itshape
        Accepted for presentation at IECON 2026, the 52nd Annual Conference of the IEEE Industrial Electronics Society.%
      }%
    }%
  }%
  \AtPageLowerLeft{%
    \raisebox{5mm}[0pt][0pt]{%
      \makebox[\paperwidth][c]{%
        \parbox[b]{0.94\paperwidth}{\centering\scriptsize
          \textcopyright{} 2026 IEEE. Personal use of this material is permitted.
          Permission from IEEE must be obtained for all other uses, in any current
          or future media, including reprinting/republishing this material for
          advertising or promotional purposes, creating new collective works, for
          resale or redistribution to servers or lists, or reuse of any copyrighted
          component of this work in other works.}%
      }%
    }%
  }%
}%
\thispagestyle{empty}
\pagestyle{empty} 

\begin{abstract}
A recently proposed method was able to generate current references with minimum stator copper loss (\acronymabsdef{SCL}{SCL}) and torque ripple over the full torque-speed range of multiphase nonsalient permanent-magnet synchronous machines (\acronymabsdef{PMSM}{PMSMs}) with nonsinusoidal back-electromotive force (\acronymabsdef{backEMF}{back-EMF}). However, it relied on large lookup tables (\acronymabsdef{LUT}{LUTs}) generated offline, requiring several hours for generation and large memory for storage. Since on-chip memory in industrial digital signal processors (\acronymabsdef{DSP}{DSPs}) is limited, this hinders practical implementation. This paper addresses this problem by analyzing how the \acronymabsref{LUT}{LUTs} of that method can be simplified and optimized, focusing on the example of a symmetrical six-phase \acronymabsref{PMSM}{PMSM} drive. Breakpoint selection (\acronymabsdef{BS}{BS}), \acronymabsref{LUT}{LUT} construction (\acronymabsdef{LC}{LC}) and interpolation approaches are evaluated to optimize the tradeoff between \acronymabsref{LUT}{LUT} simplification and performance. It is shown that the required memory and \acronymabsref{LUT}{LUT} generation time can be greatly reduced while keeping nearly the same feasible torque-speed area and performance. Simulation results confirm negligible performance degradation.
\end{abstract}

\begin{IEEEkeywords}
Lookup table, maximum torque, minimum losses, multiphase, six-phase, speed range, voltage constraint.
\end{IEEEkeywords}

\section{Introduction}
\label{sec:introduction}

Multiphase permanent-magnet synchronous machines (\acronymdef{PMSM}{PMSMs}) are well suited for applications requiring high reliability and torque density, as well as low per-phase power \cite{Taha2023TPEL}. In particular, six-phase \acronymref{PMSM}{PMSMs} offer a good tradeoff between these features and complexity \cite{Taha2023TPEL}, while symmetrical six-phase stator windings are advantageous over asymmetrical ones in terms of fault tolerance \cite{Yepes2026TIE}, dc-link capacitor size \cite{Taha2023TPEL}, etc. In addition, \acronymref{PMSM}{PMSMs} with nonsinusoidal back-electromotive force (\acronymdef{backEMF}{back-EMF}), such as those based on fractional-slot concentrated windings (\acronymdef{FSCW}{FSCWs}), can further increase torque density \cite{Wang2024TIE}. Although \acronymref{FSCW}{FSCW} \acronymref{PMSM}{PMSMs} usually exhibit negligible saliency, they also offer high fault tolerance and reduced end windings \cite{Zhang2018TAS}.

To fully exploit the torque-speed capability of \acronymref{PMSM}{PMSMs} driven by field-oriented control, suitable current references must be generated for the inner current controller according to the torque reference, actual speed and rotor position. Different current-reference generation techniques for multiphase \acronymref{PMSM}{PMSMs} with current and voltage constraints have been proposed \cite{Lu2026TPEL,Sun2010TIE,Laksar2026TIE}. However, they exhibit important drawbacks \cite{Yepes2026TIE} such as unsuitability for symmetrical six-phase \acronymref{PMSM}{PMSMs} \cite{Lu2026TPEL,Laksar2026TIE} or for any \acronymref{backEMF}{back-EMF} harmonics \cite{Lu2026TPEL,Laksar2026TIE}, inability to allow moderate torque ripple to further extend the torque-speed range \cite{Lu2026TPEL,Sun2010TIE,Laksar2026TIE}, reduced efficiency and torque-speed capability due to fewer degrees of freedom \cite{Sun2010TIE}, etc.

In \cite{Yepes2026TIE}, a nonsinusoidal current-reference generation method overcoming these limitations was introduced for nonsalient \acronymref{PMSM}{PMSMs} with any phase number, any sinusoidal or nonsinusoidal \acronymref{backEMF}{back-EMF} waveforms, and symmetrical or asymmetrical windings. It consists of an offline stage and an online stage. First, the optimum Fourier coefficients of the nonsinusoidal current references are found offline by solving a quadratic optimization problem for multiple torque-speed operating points, and stored in lookup tables (\acronymdef{LUT}{LUTs}) with uniformly spaced torque/speed breakpoints. This offline optimization minimizes stator copper loss (\acronymdef{SCL}{SCL})\footnote{In a \acronymref{PMSM}{PMSM}, \acronymref{SCL}{SCL} is commonly the most important heat source \cite{Yepes2026TIE}.} and torque ripple per operating point, while mean-torque tracking is imposed through an equality constraint. Torque ripple, peak current, and peak voltage are limited, enabling operation over the full torque-speed range. In the online stage, the current references are generated from the \acronymref{LUT}{LUTs} by linear interpolation as a function of the torque reference and actual speed \cite{Yepes2026TIE}. 

However, $4.4\MB$ was needed to store the ten required \acronymref{LUT}{LUTs}, assuming $16$ bits per value \cite{Yepes2026TIE}. This exceeds the available on-chip RAM space in most standard digital signal processors (\acronymdef{DSP}{DSPs}). Furthermore, solving the optimization problem to generate \acronymref{LUT}{LUTs} with so many entries took several hours despite using a powerful computer \cite{Yepes2026TIE}. In addition, since the rectangular \acronymref{LUT}{LUTs} contain feasible and infeasible operating points, this memory requirement of $4.4\MB$ was estimated assuming that only the values corresponding to feasible operating points were stored. However, this would produce variable-length LUT rows and require more complex indexing and access logic. Otherwise, storing the full ten rectangular \acronymref{LUT}{LUTs} from \cite{Yepes2026TIE}, with as many as $389\times790$ elements per \acronymref{LUT}{LUT}, would imply even more memory ($6.1\MB$).

This paper addresses the simplification and optimization of the \acronymref{LUT}{LUTs} for the current-reference generation method from \cite{Yepes2026TIE}, focusing on a symmetrical star-connected six-phase \acronymref{PMSM}{PMSM} drive with nonsinusoidal \acronymref{backEMF}{back-EMF}. After reviewing the method from \cite{Yepes2026TIE} (Section~\ref{sec:review}), different possibilities for simplification of its original \acronymref{LUT}{LUTs} (\acronymdef{OLUT}{OLUTs}) are introduced (Section~\ref{sec:approaches}), including various \acronymref{LUT}{LUT}-construction (\acronymdef{LC}{LC}), breakpoint-selection (\acronymdef{BS}{BS}) and interpolation approaches. Two main types of \acronymref{BS}{BS} are considered: \acronymref{LUT}{LUTs} with uniformly spaced breakpoints of various resolutions, and \acronymref{LUT}{LUTs} with nonuniform spacing obtained by an alternative \acronymref{LUT}{LUT} reduction procedure. For the selected case-study parameters and metrics (Section~\ref{sec:parameters}), the reduced \acronymref{LUT}{LUTs} are evaluated (Section~\ref{sec:uniform}) to optimize the tradeoff between \acronymref{LUT}{LUT} simplification and resulting performance. \acronymref{LUT}{LUT} simplification is assessed through generation time and memory, whereas the performance is evaluated through the reduction in the feasible torque-speed area under acceptable limits on peak line voltage, peak phase current, torque ripple and mean-torque deviation. Several options achieving a favorable compromise between simplicity and performance are identified and compared. The required memory and generation time are remarkably reduced, with a very small decrease in feasible torque-speed region. Simulation results (Section~\ref{sec:sim}) confirm negligible performance degradation, despite the remarkable \acronymref{LUT}{LUT} simplification.

\section{Current-Reference Generation Method of \cite{Yepes2026TIE}}
\label{sec:review}

\begin{figure}[t!]
\centering
\includegraphics[width=\columnwidth]{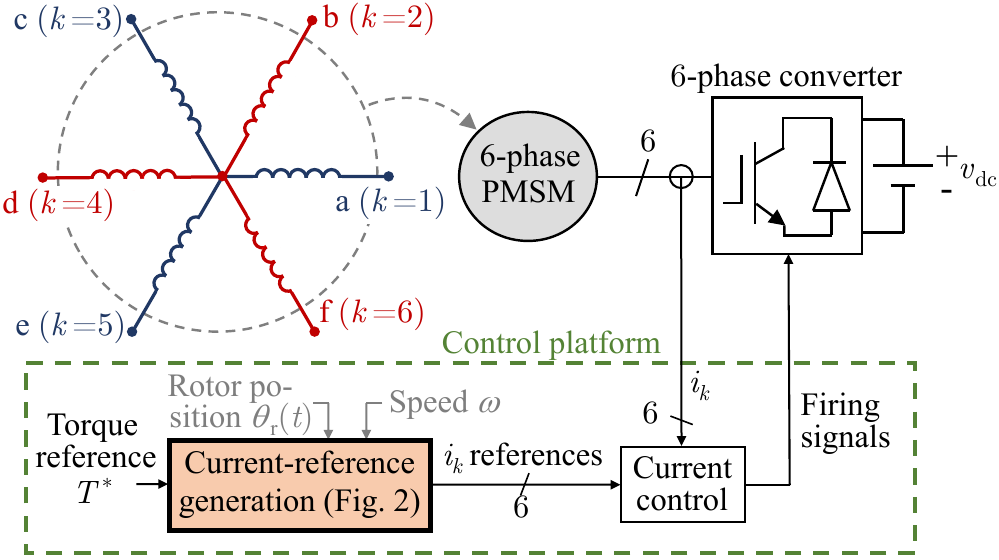}
\caption{Six-phase \acronymref{PMSM}{PMSM} drive based on field-oriented control \cite{Yepes2026TIE}, including current-reference generation, on which this paper is focused.}
\label{fig:drive}
\end{figure}

Fig.~\ref{fig:drive} depicts a \acronymref{PMSM}{PMSM} drive with field-oriented control. The current references for the inner current control are generated based on the torque $T$ reference $T^{\ast}$, electrical speed $\omega$ and rotor position $\theta_{\mathrm{r}}$.\footnote{Hereafter, for simplicity, the asterisk in $T^{\ast}$ is omitted when referring to $T^{\ast}$ breakpoints, breakpoint steps $\Delta T^{\ast}$, $T^{\ast}$-$\omega$ areas or $T^{\ast}$-$\omega$ curves.} The current reference-generation method from \cite{Yepes2026TIE} involves online and offline stages, reviewed next.

\subsection{Online Stage: Current-Reference Generation}

\begin{figure}[t!]
\centering
\includegraphics[width=0.9\columnwidth]{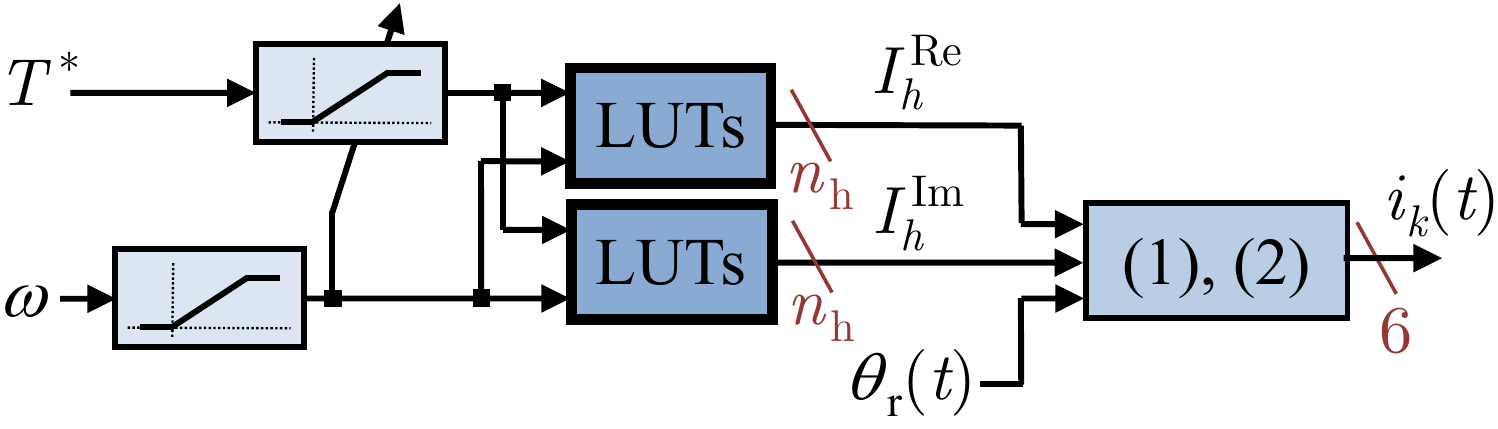}
\caption{Internal structure of the current-reference generation block in Fig.~\ref{fig:drive}, corresponding to the online stage of the method proposed in \cite{Yepes2026TIE}.}
\label{fig:online_scheme_healthy}
\end{figure}

In the online stage (see Fig.~\ref{fig:online_scheme_healthy}), the nonsinusoidal current references are generated at each instant $t$ based on the rotor electrical position $\theta_{\mathrm{r}}(t)=\omega t$ (assuming constant $\omega$ between samples) and the Fourier coefficients $I^{\mathrm{Re}}_{h}$ and $I^{\mathrm{Im}}_{h}$ of each current harmonic $h\in\left\{1,3,...,H\right\}$. The number of harmonics is $n_{h}$. $I^{\mathrm{Re}}_{h}$ and $I^{\mathrm{Im}}_{h}$ are read from $2n_{h}$ \acronymref{LUT}{LUTs} as a function of $T^{\ast}$ and $\omega$, limited (see the saturation blocks) to their feasible ranges. In particular, for a symmetrical six-phase \acronymref{PMSM}{PMSM}, the current reference $i_{k}(t)$ of each phase $k\in\left\{1,2,...,6\right\}$ is \cite{Yepes2026TIE}
\begin{equation}
\label{ec:online}
i_{k}(t)=\sum_{h}I^{\mathrm{Re}}_{h}\cos\left(h\theta_{k}(t)\right)+\sum_{h}I^{\mathrm{Im}}_{h}\sin\left(h\theta_{k}(t)\right)\vspace{-8pt}
\end{equation} 
where 
\begin{equation}\label{ec:theta_k}
\theta_{k}(t)=\theta_{\mathrm{r}}(t)-\left(k-1\right)\frac{2\pi}{6}.
\end{equation}

\subsection{Offline Stage: \texorpdfstring{\acronymref{OLUT}{OLUT}}{OLUT} Generation}
\label{sec:offline}

In the offline stage, the $2n_{h}$ \acronymref{LUT}{LUTs} of $I^{\mathrm{Re}}_{h}$ and $I^{\mathrm{Im}}_{h}$ in Fig.~\ref{fig:online_scheme_healthy} are generated by solving an optimization problem for each operating point ($T^{\ast},\omega$), for subsequent storage in a \acronymref{DSP}{DSP} \cite{Yepes2026TIE}. These \acronymref{LUT}{LUTs} are in principle rectangular, but only a tapered ($T^{\ast},\omega$) region is feasible and of interest \cite{Yepes2026TIE}. To sweep the feasible operating points, $T^{\ast}$ is increased from zero toward positive $T^{\ast}$ with uniform steps $\Delta T$, and for each $T^{\ast}$, $\omega$ is increased from zero toward positive $\omega$ with uniform steps $\Delta\omega$, until no solution is found. The process is then repeated for $T^{\ast}$ increasing from zero toward negative $T^{\ast}$. Only $\omega\geq0$ is considered, since the outputs for $\omega<0$ would be very similar to those of $\omega>0$ with opposite $T^{\ast}$ sign.  

The optimization problem per ($T^{\ast},\omega$) minimizes torque ripple with strict priority and, among the solutions with minimum torque ripple, the \acronymref{SCL}{SCL}. This is done while imposing the equality constraint that the mean permanent-magnet torque $\overline{T}$ equals that of the reference ($\overline{T}=T^{\ast}$), as well as inequality constraints that limit the peak line voltage $v_{\mathrm{pk}}$, peak phase current $i_{\mathrm{pk}}$ and torque ripple $\tau$ to the respective limits\footnote{The limit  $\tau^{\mathrm{MX}}$ from \cite{Yepes2026TIE} is here called $\tau^{\mathrm{mx}}$ for the sake of convenience.} \cite{Yepes2026TIE}:
\begin{equation}
\label{ec:constraints}
\overline{T}_{\mathrm{err}}=0;\  \  \  \  v_{{\mathrm{pk}}}\leq v^{\mathrm{mx}}_{{\mathrm{pk}}};\  \  \  \  i_{{\mathrm{pk}}}\leq i^{\mathrm{mx}}_{{\mathrm{pk}}};\  \  \  \  \tau\leq\tau^{\mathrm{mx}}
\end{equation}
where $\overline{T}_{\mathrm{err}}=\overline{T}-\overline{T}^{\ast}$ is mean $T$ error. This is solved using MATLAB \texttt{quadprog} with negligible numerical tolerances.

The $\omega$ breakpoints below $\omega^{\downarrow}$ are removed, as the \acronymref{LUT}{LUT} values there do not vary with $\omega$ (the $v_{{\mathrm{pk}}}$ constraint is inactive) \cite{Yepes2026TIE}.

\section{Definition of \texorpdfstring{\acronymref{LUT}{LUT}}{LUT} Simplification Approaches}
\label{sec:approaches}

Different \acronymref{LC}{LC}, \acronymref{BS}{BS} and interpolation approaches for \acronymref{LUT}{LUT} reduction are considered.

The reduced \acronymref{LUT}{LUTs} may be constructed using \approachref{LC1} or \approachref{LC2}. 
\begin{itemize}[leftmargin=1em,itemsep=0pt,topsep=0pt,parsep=0pt,partopsep=0pt]
    \item[\raisebox{0.5ex}{\tiny$\bullet$}] \textit{\approachdef{LC1} (downsampling):} High-resolution OLUTs according to \cite{Yepes2026TIE} using small steps $\Delta T$ and $\Delta\omega$ are first generated. Reduced LUTs are then obtained by retaining only OLUT entries with steps $\lambda\Delta T$ and $\lambda\Delta\omega$, where $\lambda$ is an integer.
		\item[\raisebox{0.5ex}{\tiny$\bullet$}] \textit{\approachdef{LC2} (direct generation):} High-resolution OLUTs are not generated. Instead, the offline LUT-generation procedure from \cite{Yepes2026TIE} (Section~\ref{sec:offline}) is run at a reduced breakpoint set, using steps $\lambda\Delta T$ and $\lambda\Delta\omega$, to directly form the new \acronymref{LUT}{LUTs}.	
\end{itemize}\vspace{0.4ex}
\approachref{LC1} requires prior \acronymref{OLUT}{OLUT} generation, whereas \approachref{LC2} does not.

\begin{figure}[t]
\centering
\footnotesize
\resizebox{0.86\columnwidth}{!}{%
\begin{tikzpicture}[
    node distance=3.2mm and 5mm,
    >=Latex,
    block/.style={
        rectangle,
        rounded corners,
        draw,
        align=center,
        minimum width=3.0cm,
        minimum height=6.2mm,
        inner sep=2.5pt
    },
    blocksmall/.style={
        rectangle,
        rounded corners,
        draw,
        align=center,
        minimum width=1.8cm,
        minimum height=4.6mm,
        inner sep=2pt
    },
    decision/.style={
        diamond,
        draw,
        align=center,
        aspect=6.0,
        inner sep=2pt
    },
    line/.style={draw, -Latex}
]

\node[blocksmall] (start) {Load \acronymref{OLUT}{OLUTs}};
\node[block, below=of start] (load) {Set target \acronymref{LUT}{LUT} size, limits ($v^{\mathrm{mx}}_{\mathrm{pk}}$, $i^{\mathrm{mx}}_{\mathrm{pk}}$ and $\tau^{\mathrm{mx}}$),\\ allowed excess $\epsilon$, evaluation grid, etc.};
\node[block, below=of load] (preevaluate) {Evaluate removing each of the \acronymref{LUT}{LUT} rows and columns:\\ feasible-area loss \& error metric $\delta$ in (\ref{ec:delta})};
\node[block, below=of preevaluate] (evaluate) {For the evaluation-grid points affected by the last breakpoint\\ removal, update evaluation of removing each of the remaining\\ \acronymref{LUT}{LUT} rows and columns: feasible-area loss \& error metric $\delta$ in (\ref{ec:delta})};
\node[block, below=of evaluate] (select) {Select candidate row/column with minimum feasible-area loss\\ (tie-break: minimum error $\delta$)};
\node[block, below=of select] (update) {Update retained breakpoint set and feasible-region map};
\node[decision, below=2.5mm of update] (stop) {Target \acronymref{LUT}{LUT} size reached?};
\node[blocksmall, below=2.5mm of stop] (end) {Store final breakpoint sets and reduced \acronymref{LUT}{LUTs}};

\draw[line] (start) -- (load);
\draw[line] (load) -- (preevaluate);
\draw[line] (preevaluate.west) -- ++(-9mm,0) |- (select.west);
\draw[line] (evaluate) -- (select);
\draw[line] (select) -- (update);
\draw[line] (update) -- (stop);
\draw[line] (stop) -- node[pos=0.4,right] {Yes} (end);

\draw[line] (stop.east) -- ++(1.8,0) node[midway, above] {No}
                  |- (evaluate.east);

\end{tikzpicture}}
\caption{Flowchart for \acronymref{LUT}{LUT} reduction with nonuniform spacing, i.e., \approachref{BS4}.}
\label{fig:flowchart_LUT_reduction}
\end{figure}
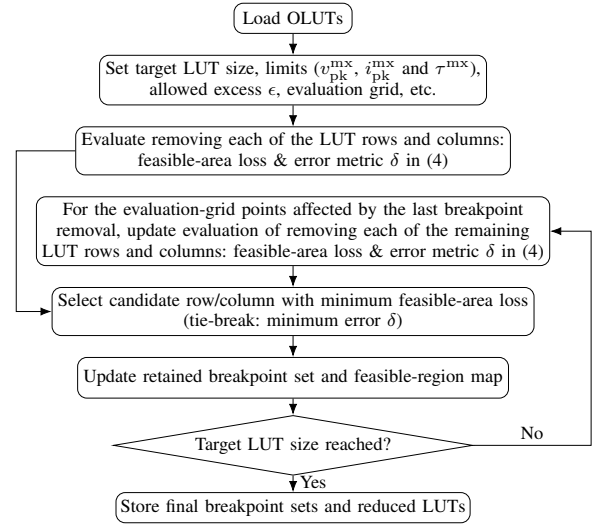

To select the breakpoints, several \acronymref{BS}{BS} approaches are defined next. \approachref{BS1}--\approachref{BS3} and \approachreftext{BS3}{BS3B} use (mostly) uniform spacing, whereas \approachref{BS4} uses nonuniform spacing. \approachref{BS1}, \approachref{BS2} and \approachref{BS4} rely on \approachref{LC1}, whereas \approachref{BS3} and \approachreftext{BS3}{BS3B} rely on \approachref{LC2}. Fig.~\ref{fig:flowchart_LUT_reduction} depicts the \approachref{BS4} flowchart, and the grids in Fig.~\ref{fig:plotting_area} indicate the \acronymref{BS}{BSs} (starting breakpoints, in red; retained \acronymref{OLUT}{OLUT} outermost breakpoints, by bold tick labels) for several examples together with the \acronymref{OLUT}{OLUT} $T$-$\omega$ limit curve in purple. The other curves in Fig.~\ref{fig:plotting_area}, discussed later, correspond to the reduced \acronymref{LUT}{LUTs}.
\begin{itemize}[leftmargin=1em,itemsep=0pt,topsep=0pt,parsep=0pt,partopsep=0pt]
    \item[\raisebox{0.5ex}{\tiny$\bullet$}] \textit{\approachdef{BS1} (uniform top-down $\Delta T$):} As indicated by the grid in Fig.~\ref{fig:PlotArea_top2bottom}, using \approachref{LC1}, the $T$ breakpoints are selected from the \acronymref{OLUT}{OLUT} highest positive one (retained) using equal decrements $\lambda \Delta T$, except for a smaller last step to also retain the lowest negative \acronymref{OLUT}{OLUT} $T$ breakpoint. The \acronymref{OLUT}{OLUT} highest $\omega$ breakpoint is retained, and the other $\omega$ breakpoints are set using equal decrements $\lambda\Delta \omega$ until $\omega^{\downarrow}$ ($796\rpm$), below which the \acronymref{OLUT}{OLUT} values do not vary with $\omega$ \cite{Yepes2026TIE}. 
    \item[\raisebox{0.5ex}{\tiny$\bullet$}] \textit{\approachdef{BS2} (uniform zero-based $\Delta T$):} As shown in Fig.~\ref{fig:PlotArea_zero2extremes}, using \approachref{LC1}, the $\omega$ breakpoints are set as in \approachref{BS1}. The $T$ breakpoints are defined starting from zero torque and applying $\lambda\Delta T$ toward the largest positive and negative feasible torques of the \acronymref{OLUT}{OLUTs} (also retained as breakpoints), leaving the outermost interval on each side smaller than $\lambda\Delta T$.
    \item[\raisebox{0.5ex}{\tiny$\bullet$}] \textit{\approachdef{BS3} (uniform zero-based $\Delta T\&\Delta \omega$):} As shown in Figs.~\ref{fig:PlotArea_zero2extremes_zerow2top_lambda17} and~\ref{fig:PlotArea_zero2extremes_zerow2top}, using \approachref{LC2}, uniform breakpoint spacings $\lambda\Delta T$ and $\lambda\Delta\omega$ are applied starting from zero torque (toward positive and negative $T$) and speed (toward positive $\omega$) until no solution of the optimization problem is found. The $\omega$ breakpoints below $\omega^{\downarrow}$ are removed. The outermost $\omega$ and $T$ breakpoints from the high-resolution \acronymref{OLUT}{OLUTs} are not included (may be unknown, if such \acronymref{OLUT}{OLUTs} are not previously generated). A variant of this approach, denoted as \approachreftext{BS3}{BS3B} and shown in Fig.~\ref{fig:PlotArea_3B}, then applies high-resolution steps with smaller $\lambda$ (e.g., $\lambda=1$) from the outermost breakpoints for which a solution was found, until new outermost feasible breakpoints are identified and included in the \acronymref{BS}{BS}. These would match the outermost breakpoints of the high-resolution \acronymref{OLUT}{OLUTs}, but saving the computation time that would be required to generate the OLUTs beforehand.
    \item[\raisebox{0.5ex}{\tiny$\bullet$}] \textit{\approachdef{BS4} (nonuniform):} A nonuniform \acronymref{BS}{BS} [see Fig.~\ref{fig:PlotArea_1209}] is obtained by the \approachref{LC1}-based flowchart in Fig.~\ref{fig:flowchart_LUT_reduction}. Selected $T$ and $\omega$ breakpoints of the \acronymref{OLUT}{OLUTs} are iteratively removed one by one, until the total number of \acronymref{LUT}{LUT} entries is below a specified target. A high-resolution grid with uniform $T$ and $\omega$ spacing is defined beforehand for evaluation, using linear interpolation between the remaining entries. The feasible $T$-$\omega$ region is the ($T^{\ast},\omega$) set for which the variables of interest remain within admissible bounds allowing an excess $\epsilon$ over the nominal limits (further detailed in the next section). At each iteration, the removed breakpoint is the one causing the smallest reduction in feasible $T$-$\omega$ area or, in case of a tie, the one yielding the lowest value of the error metric
\begin{equation}\label{ec:delta}
\delta=\frac{1}{N_{\mathrm{points}}}\sum_{\mathrm{points}}
\left[\left(\overline{T}-\overline{T}_{\mathrm{OLUT}}\right)^{2}
+\left(\tau-\tau_{\raisebox{-0.25ex}{$\scriptstyle \mathrm{OLUT}$}}  \right)^{2}\right]
\end{equation}
where the summation is performed over all evaluation-grid points affected by the evaluated breakpoint removal (i.e., those in the feasible area between the previous breakpoint and the following one in the same dimension); $N_{\mathrm{points}}$ is the number of such points; and the $\mathrm{OLUT}$ subscript denotes the $\overline{T}$ and $\tau$ values obtained for the \acronymref{OLUT}{OLUT} at the same evaluation-grid point. Note that $\overline{T}_{\mathrm{OLUT}}\approx T^{\ast}$. For this evaluation in Fig.~\ref{fig:flowchart_LUT_reduction}, double-precision (64-bit) floating-point format is used to ensure \acronymref{BS}{BS} accuracy, but lower precision may be used for resulting-\acronymref{LUT}{LUT} storage.
\end{itemize}\vspace{0.4ex}

\afterpage{%
  \clearpage
  \begin{figure*}[!p]
\centering
\subfloat[\label{fig:PlotArea_top2bottom}]{\includegraphics[width=\columnwidth]{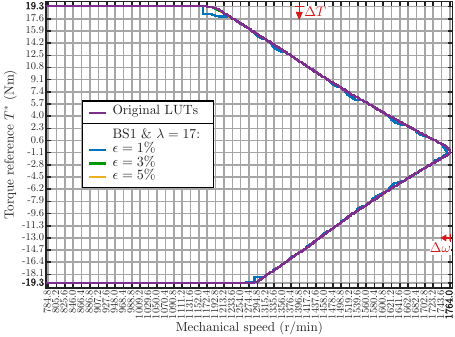}}\quad
\subfloat[\label{fig:PlotArea_zero2extremes}]{\includegraphics[width=\columnwidth]{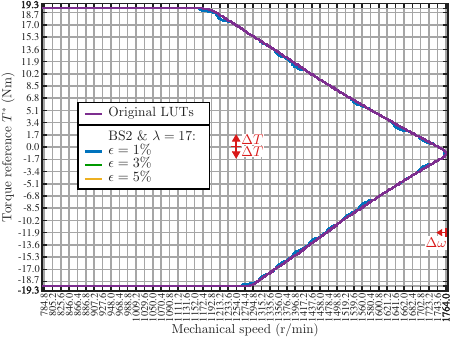}}\\
\subfloat[\label{fig:PlotArea_zero2extremes_zerow2top_lambda17}]{\includegraphics[width=\columnwidth]{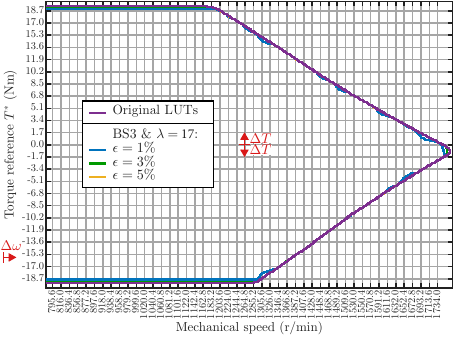}}\quad
\subfloat[\label{fig:PlotArea_zero2extremes_zerow2top}]{\includegraphics[width=\columnwidth]{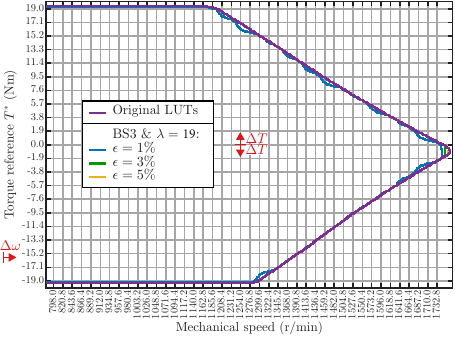}}\\
\subfloat[\label{fig:PlotArea_3B}]{\includegraphics[width=\columnwidth]{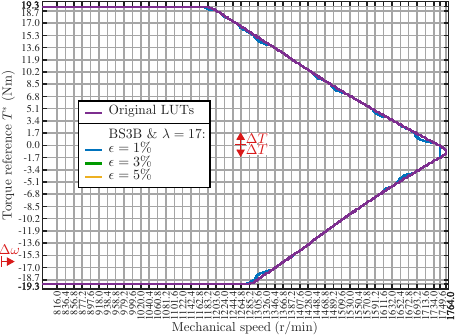}}\quad
\subfloat[\label{fig:PlotArea_1209}]{\includegraphics[width=\columnwidth]{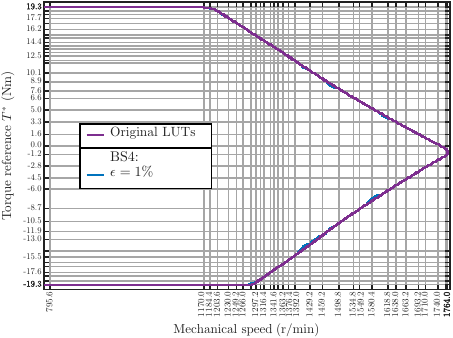}}
\caption{Grid indicating $T$ and $\omega_{\mathrm{m}}$ \acronymref{BS}{BSs} of reduced \acronymref{LUT}{LUTs} (starting \acronymref{BS}{BS}, in red; retained \acronymref{OLUT}{OLUT} outermost breakpoints, by bold tick labels), and limit curves of the feasible area for \acronymref{OLUT}{OLUTs} \cite{Yepes2026TIE} and for reduced \acronymref{LUT}{LUTs}, evaluated using linear interpolation and different $\epsilon$. For speeds below those shown, the limit boundaries continue horizontally. Green and yellow curves practically overlap with purple ones, not with blue ones. (a)~\approachref{BS1} and $\lambda=17$. (b)~\approachref{BS2} and $\lambda=17$, (c)~\approachref{BS3} and $\lambda=17$. (d)~\approachref{BS3} and $\lambda=19$. (e)~\approachreftext{BS3}{BS3B} and $\lambda=17$. (f)~\approachref{BS4} and $\epsilon=1\%$, omitting some tick labels (cf. Table~\ref{tab:nonuniform_breakpoints}) to avoid visual clutter.}
\label{fig:plotting_area}
\end{figure*}
  \clearpage
}

For each $T$-$\omega$ breakpoint pair such that no feasible solution is found when solving the optimization problem of Section~\ref{sec:offline} (when generating the \acronymref{OLUT}{OLUTs} for \approachref{LC1}, or the reduced \acronymref{LUT}{LUTs} for \approachref{LC2}) for such point, the values in the reduced \acronymref{LUT}{LUT} are set equal to those of the highest $\omega$ breakpoint for which a solution was found for the same $T$. 

When evaluating any of the reduced \acronymref{LUT}{LUTs}, different types of interpolation may also be adopted, e.g., those available in the MATLAB \texttt{griddedInterpolant} function \cite{MathWorks_griddedInterpolant}: linear, Modified Akima, cubic convolution, and cubic spline.

\section{Parameters and Metrics for the Evaluation}
\label{sec:parameters}

\subsection{Parameters of \texorpdfstring{\acronymref{PMSM}{PMSM}}{PMSM} and of \texorpdfstring{\acronymref{LUT}{LUT}}{LUT} Generation Algorithm}

\begin{figure}[t!]
\centering
\includegraphics[width=\columnwidth]{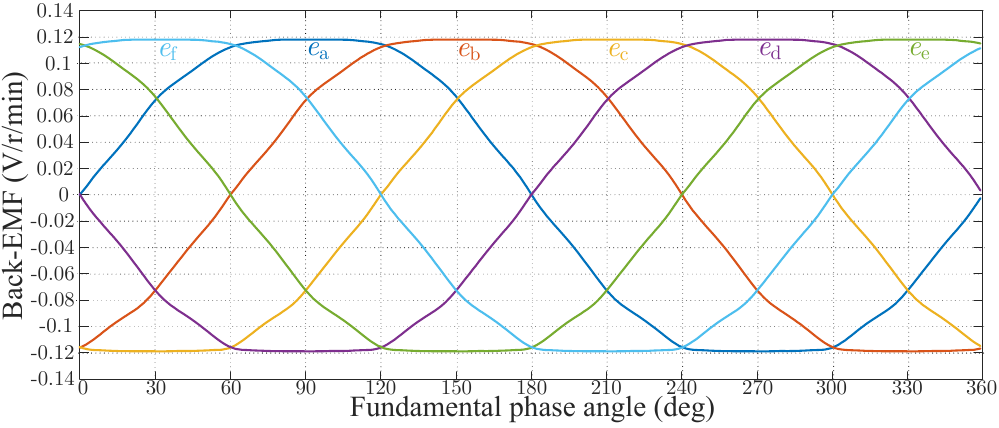}
\caption{\acronymref{PMSM}{PMSM} \acronymref{backEMF}{back-EMF} waveforms normalized by mechanical speed \cite{Yepes2026TIE}.}
\label{fig:back_EMF}
\end{figure}

For the subsequent analysis, the \acronymref{PMSM}{PMSM} considered in this paper is the same as in \cite{Yepes2026TIE}. Namely, it is a six-phase nonsalient 12-slot/10-pole \acronymref{PMSM}{PMSM} with symmetrical windings, using star connection with a single neutral point. The \acronymref{PMSM}{PMSM} ratings for current, voltage, torque and speed are $3.2\A$, $100\V$, $16.7\Nm$ and $1100\rpm$. Using the vector space decomposition (\acronymdef{VSD}{VSD}) \cite{Yepes2018TPEL}, the stator inductance is $12\mH$, $11.3\mH$ and $9.4\mH$ in the $\alpha_{1}\beta_{1}$ plane,  $\alpha_{2}\beta_{2}$ plane, and zero-sequence $\alpha_{3}$ axis, respectively. The stator resistance is $1.4\ohm$ in all \acronymref{VSD}{VSD} subspaces. The \acronymref{backEMF}{back-EMF} waveforms are shown in Fig.~\ref{fig:back_EMF}. Saliency, cogging torque and flux saturation are neglected \cite{Yepes2026TIE}.

Most parameters of the \acronymref{LUT}{LUT} generation algorithm are here set as in \cite{Yepes2026TIE}: $v^{\mathrm{mx}}_{{\mathrm{pk}}}=290\V$, $i^{\mathrm{mx}}_{{\mathrm{pk}}}=4.34\A$, $\tau^{\mathrm{mx}}=2\Nm$, and $h\in\{1,3,5,7,9\}$ ($H=9$, $n_{h}=5$). As shown in \cite{Yepes2026TIE}, $H=9$ preserves nearly the same torque-speed range as higher $H$, while avoiding extra complexity. The number of samples per period $n_{\mathrm{t}}$ is set to $150$ for high accuracy \cite{Yepes2026TIE}. 

\subsection{Evaluation Metrics}
\label{ec:metrics}

The reduced \acronymref{LUT}{LUTs} will be evaluated in terms of \acronymref{LUT}{LUT} simplification (required generation time and memory) and performance (feasible $T$-$\omega$ area).

For memory assessment, $16$ bits per \acronymref{LUT}{LUT} entry are assumed \cite{Yepes2026TIE}, and the full rectangular \acronymref{LUT}{LUTs}, including infeasible operating points, are stored for indexing simplicity. For the offline \acronymref{LUT}{LUT} generation time, a computer with $5200$-MHz DDR5 RAM and i9-14900K processor is employed \cite{Yepes2026TIE}.

For each \acronymref{LUT}{LUT} set, the feasible $T$-$\omega$ range is defined as the $T$-$\omega$ area where variables of interest satisfy an allowed excess $\epsilon$. For $\epsilon$ in p.u., the allowed values within this area are $v_{{\mathrm{pk}}}\leq(1+\epsilon)v^{\mathrm{mx}}_{{\mathrm{pk}}}$, $i_{{\mathrm{pk}}}\leq(1+\epsilon)i^{\mathrm{mx}}_{{\mathrm{pk}}}$ and $\tau\leq(1+\epsilon)\tau^{\mathrm{mx}}$. Compared with (\ref{ec:constraints}), $\epsilon$ represents allowed excess in $v^{\mathrm{mx}}_{{\mathrm{pk}}}$, $i^{\mathrm{mx}}_{{\mathrm{pk}}}$ and $\tau^{\mathrm{mx}}$.\footnote{$\epsilon>0$ need not imply above-rated operation, since $v^{\mathrm{mx}}_{{\mathrm{pk}}}$, $i^{\mathrm{mx}}_{{\mathrm{pk}}}$ and $\tau^{\mathrm{mx}}$ may include margins below rated values \cite{Yepes2026TIE}.} In addition, the mean-torque error $\overline{T}_{\mathrm{err}}=\overline{T}-\overline{T}^{\ast}$ must satisfy $\left|\overline{T}_{\mathrm{err}}\right|\leq\epsilon T^{\mathrm{rt}}$, where $T^{\mathrm{rt}}$ is the \acronymref{PMSM}{PMSM} rated torque. The $\tau$ excess $\tau_{\mathrm{err}}$ over that of the \acronymref{OLUT}{OLUTs} must satisfy $\tau_{\mathrm{err}}=\tau-\tau_{\raisebox{-0.25ex}{$\scriptstyle \mathrm{OLUT}$}}\leq\epsilon T^{\mathrm{rt}}$. The feasible $T$-$\omega$ area for a certain $\epsilon$ is assessed by evaluating\footnote{The allowed excess $\epsilon$ during the evaluation of the feasible $T$-$\omega$ area of given \acronymref{LUT}{LUTs} is independent of the numerical tolerances (negligible) used with \texttt{quadprog} to solve the quadratic optimization (Section~\ref{sec:offline}) during the generation of the \acronymref{OLUT}{OLUTs} for \approachref{LC1} or of the reduced \acronymref{LUT}{LUTs} for \approachref{LC2}.} $v_{{\mathrm{pk}}}$, $i_{{\mathrm{pk}}}$, $\tau$, $\overline{T}_{\mathrm{err}}$ and $\tau_{\mathrm{err}}$ over a uniform evaluation grid with resolution $0.1\Nm$ and $1.2\rpm$ for $T$ and $\omega$, respectively, which also match the resolution $\Delta T$ and $\Delta \omega_{\mathrm{m}}$ of the \acronymref{OLUT}{OLUTs}, with $\omega_{\mathrm{m}}$ being mechanical speed. For evaluation-grid points located between (outside) \acronymref{LUT}{LUT} breakpoints, interpolation (extrapolation) is used. Only points within the \acronymref{OLUT}{OLUT} feasible region are evaluated. For $\omega$ below the lowest $\omega$ breakpoint ($\approx\omega^{\downarrow}$), the \acronymref{LUT}{LUT} outputs of that breakpoint are taken without extrapolation \cite{Yepes2026TIE}.

\begin{table*}[t!]
    \centering\footnotesize
    \caption{Comparison of Selected Reduced \acronymref{LUT}{LUTs}}
    \label{tab:metrics}
    \setlength{\tabcolsep}{3.8pt}
    \begin{tabular}{cccccccccc}
			\toprule
        \thead{\acronymref{BS}{BS}} & \thead{\acronymref{LC}{LC}} &
        \thead{Downsampling\\factor\\$\lambda$} &
        \thead{Resolution\\$\lambda\Delta T $$\times$$ \lambda\Delta\omega_{\mathrm{m}}$\\(Nm$\times$r/min)} &
        \thead{Stored\\ $T$$\times$$\omega$ entries\\per \acronymref{LUT}{LUT}} &
        \thead{Feasible\\$T$-$\omega$ area (\%)\\for $\epsilon=1\%$} &
        \thead{Area-size\\tradeoff vs.\\alternatives} &
        \thead{$T$-$\omega_{\mathrm{m}}$\\limit curve} &
        \thead{Total\\memory\\(MB)} &
        \thead{Total \acronymref{LUT}{LUT}\\generation\\time (min)} \\
        \midrule
        \acronymref{OLUT}{OLUTs} \cite{Yepes2026TIE}  & \approachref{LC2} & $1$ & $0.1$$\times$$1.2$  & $389$$\times$$790\approx3$$\cdot$$10^{5}$  & $100$  & Fig.~\ref{fig:areaVSsize} (stars) & Fig.~\ref{fig:plotting_area} (purple) & $6.15$  & $574$ \\
        \approachref{BS1} & \approachref{LC1} & $17$ & $1.7$$\times$$20.4$ & $24$$\times$$49=1176$  & $99.1$ & Fig.~\ref{fig:areaVSsize_top2bottom} (diamonds) & Fig.~\ref{fig:PlotArea_top2bottom} & $0.024$ & $574$ \\
        \approachref{BS2} & \approachref{LC1} & $17$ & $1.7$$\times$$20.4$$^{\ast}$ & $25$$\times$$49=1225$  & $99.4$ & Fig.~\ref{fig:areaVSsize_zero2extremes} (diamonds) & Fig.~\ref{fig:PlotArea_zero2extremes} & $0.025$ & $574$ \\
        \approachref{BS3} & \approachref{LC2} & $17$ & $1.7$$\times$$20.4$$^{\ast}$ & $23$$\times$$47=1081$  & $97.2$ & Fig.~\ref{fig:areaVSsize_zero2extremes_zerow2top} (diamonds) & Fig.~\ref{fig:PlotArea_zero2extremes_zerow2top_lambda17} & $0.022$ & $9$ \\
        \approachref{BS3} & \approachref{LC2} & $19$ & $1.9$$\times$$22.8$ & $21$$\times$$42=882$   & $98.3$ & Fig.~\ref{fig:areaVSsize_zero2extremes_zerow2top} (triangles) & Fig.~\ref{fig:PlotArea_zero2extremes_zerow2top} & $0.018$ & $9$ \\
				\approachreftext{BS3}{BS3B} & \approachref{LC2} & $17$ & $1.7$$\times$$20.4$$^{\ast}$ & $25$$\times$$48=1200$ & $99.3$ & $-$ & Fig.~\ref{fig:PlotArea_3B} & $0.024$ & $10$ \\
        \approachref{BS4} & \approachref{LC1} & $-$  & Nonuniform        & $39$$\times$$31=1209$  & $99.8$ & Fig.~\ref{fig:areaVSsize_approach4} (square) & Fig.~\ref{fig:PlotArea_1209} & $0.024$ & $1549$ \\
        \bottomrule
    \end{tabular}
\begin{tablenotes}[flushleft]
\item{$^{\ast}$As an exception, one, two or three of the outermost torque/speed breakpoints are separated with a smaller spacing.}
\end{tablenotes}
\end{table*}

\begin{table}[t!]
    \centering\footnotesize
    \caption{Feasible $T$-$\omega$ Area Relative to Original \cite{Yepes2026TIE} for \approachref{BS2} Using $\lambda=17$ and $\epsilon=1\%$ Depending on Interpolation Method}
		\label{tab:interpolation_method}
    \setlength{\tabcolsep}{15pt}
    \begin{tabular}{lc}
        \toprule\\[-10pt]
         Interpolation method  & Feasible $T$-$\omega$ area ($\%$) \\
        \midrule
				 Linear (selected) & $99.4$ \\  
			Modified Akima & $93.2$ \\ 
			Cubic convolution & $84.3$ \\
			Cubic spline & $84.3$ \\
        \bottomrule
    \end{tabular}
\end{table}

\begin{table*}[t!]
    \centering\footnotesize
    \caption{Nonuniformly Spaced Breakpoints After Applying \approachref{BS4} With Target \acronymref{LUT}{LUT} Size of 1225 Entries}
    \label{tab:nonuniform_breakpoints}
    \renewcommand{\arraystretch}{1.1}
    \setlength{\tabcolsep}{0pt}
    \begin{tabular*}{\textwidth}{@{}>{\centering\arraybackslash}p{0.15\textwidth}@{\hspace{0.01\textwidth}}>{\centering\arraybackslash}p{0.84\textwidth}@{}}
        \toprule
        \shortstack[c]{Speed breakpoints\\$(\mathrm{r/min})$}
        &
        \shortstack[c]{$796,\,1170,\,1184,\,1204,\,1230,\,1249,\,1266,\,1285,\,1297,\,1308,\,1316,\,1332,\,1342,\,1350,\,1363,\,1376,$\\[2pt]
        $1392,\,1429,\,1459,\,1499,\,1535,\,1549,\,1580,\,1619,\,1638,\,1663,\,1693,\,1710,\,1740,\,1760,\,1764$}
        \\
        \addlinespace[5pt]

        \shortstack[c]{Torque breakpoints\\$(\mathrm{Nm})$}
        &
        \shortstack[c]{$-19.3,\,-18.8,\,-18.1,\,-17.6,\,-16.8,\,-16.2,\,-15.5,\,-14.6,\,-13.0,\,-11.9,\,-10.5,\,-8.7,\,-6.0,\,-4.5,\,-2.8,\,-1.2,\,0.0,$\\[2pt]
        $1.6,\,3.3,\,5.0,\,6.6,\,7.6,\,8.9,\,10.1,\,11.5,\,11.9,\,12.5,\,13.0,\,13.5,\,13.9,\,14.4,\,15.0,\,15.4,\,16.2,\,17.0,\,17.7,\,18.3,\,18.8,\,19.3$}
        \\
        \bottomrule
    \end{tabular*}
\end{table*}

\begin{figure}[!t]
\centering
\subfloat[\label{fig:areaVSsize_top2bottom}]{\includegraphics[width=0.87\linewidth]{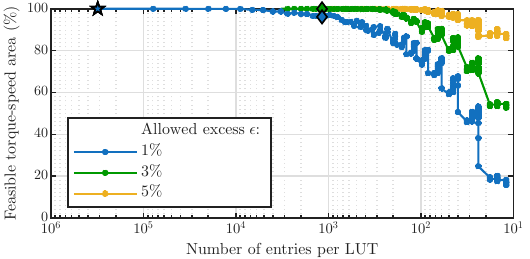}}\\
\subfloat[\label{fig:areaVSsize_zero2extremes}]{\includegraphics[width=0.87\linewidth]{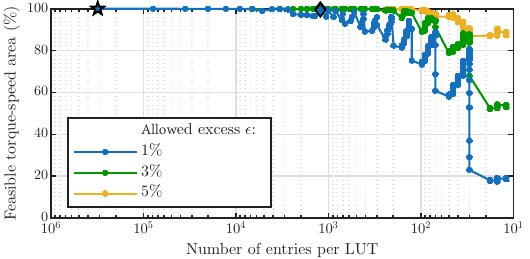}}\\
\subfloat[\label{fig:areaVSsize_zero2extremes_zerow2top}]{\includegraphics[width=0.87\linewidth]{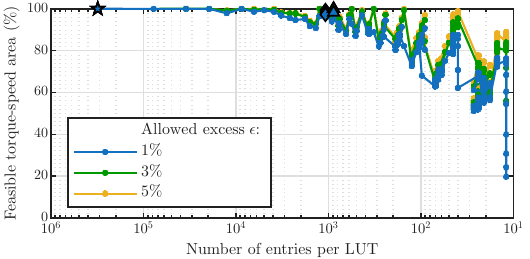}}\\
\subfloat[\label{fig:areaVSsize_approach4}]{\includegraphics[width=0.87\linewidth]{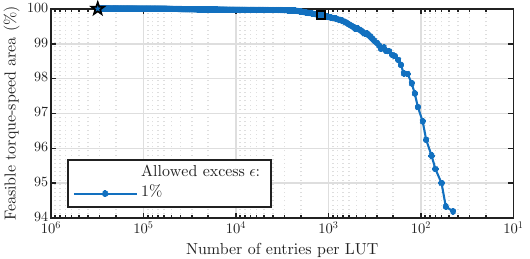}}
\caption{Feasible $T$-$\omega$ area, normalized by that of the \acronymref{OLUT}{OLUTs} \cite{Yepes2026TIE}, versus the number of entries per \acronymref{LUT}{LUT}, evaluated using various $\epsilon$ (colors) and linear interpolation between \acronymref{LUT}{LUT} breakpoints. Stars indicate $\lambda=1$ (i.e., \acronymref{OLUT}{OLUTs}); diamonds, $\lambda=17$; triangles, $\lambda=19$; and a square, $1209$ entries obtained by \approachref{BS4} with target \acronymref{LUT}{LUT} size of $1225$. (a)~\approachref{BS1}. (b)~\approachref{BS2}.  (c)~\approachref{BS3}.  (d)~\approachref{BS4}.}
\label{fig:areaVSsize}
\end{figure}

\section{Study of \texorpdfstring{\acronymref{LUT}{LUT}}{LUT} Simplification and Performance}
\label{sec:uniform}

The reduced \acronymref{LUT}{LUTs} are evaluated for the \acronymref{BS}{BS}, \acronymref{LC}{LC} and interpolation approaches according to the introduced metrics.

Fig.~\ref{fig:areaVSsize} shows the feasible $T$-$\omega$ area (for $\epsilon = 1\%,\,3\%,$ and $5\%$) versus the number of entries per \acronymref{LUT}{LUT}, for $\lambda$ swept from $1$ to $300$ for \approachref{BS1} and \approachref{BS2}, and from $1$ to $193$ for \approachref{BS3}. While the \acronymref{LUT}{LUT} size (resolution) varies, the resolution of the evaluation grid is kept as aforesaid. Linear interpolation (extrapolation) is adopted.
 The feasible $T$-$\omega$ area is normalized by that of the \acronymref{OLUT}{OLUTs}, evaluated on the same dense grid and under the same $\epsilon$. Fig.~\ref{fig:areaVSsize} was obtained by using just $16$ bits per \acronymref{LUT}{LUT} entry (as for memory storage): $16$-bit signed fixed-point values allocated as $8$ bits for the integer part and $8$ bits for the fractional part. It was verified that the use of $64$-bit instead of $16$-bit \acronymref{LUT}{LUT} entries produces no noticeable differences.

Fig.~\ref{fig:areaVSsize} shows that the \acronymref{LUT}{LUT} size can be greatly reduced with little performance loss. Using \approachref{BS1} with $\lambda=17$ [diamonds in Fig.~\ref{fig:areaVSsize_top2bottom}], each \acronymref{LUT}{LUT} is reduced from $389\times790\approx3\cdot10^{5}$ entries (stars) to $24\times49=1176$, i.e., by a factor of $261$. For these points, the feasible $T$-$\omega$ area decreases by only $0.9\%$, $0.0\%$ and $0.0\%$ for $\epsilon=1\%$, $3\%$ and $5\%$, respectively. This agrees with Fig.~\ref{fig:PlotArea_top2bottom}, where the limit curves for $\epsilon=3\%$ and $5\%$ practically overlap with the \acronymref{OLUT}{OLUT} limit (purple). For $\epsilon=1\%$, the little reduction ($0.9\%$) occurs in Fig.~\ref{fig:PlotArea_top2bottom} mainly between the two highest $T$ breakpoints. For the same $\lambda=17$ but using \approachref{BS2}, Fig.~\ref{fig:PlotArea_zero2extremes} shows that this approach, by adding an extra $T$ breakpoint in that region, is very effective in mitigating the feasible-area reduction for $\epsilon=1\%$ near the highest $T$, compared with Fig.~\ref{fig:PlotArea_top2bottom}. In fact, this increases the feasible area for $\lambda=17$ and $\epsilon=1\%$ from $99.1\%$ to $99.4\%$ [blue diamonds in Fig.~\ref{fig:areaVSsize_top2bottom} and~(b)], just with an extra $T$ breakpoint, giving $25\times49=1225$ entries. The main features of these two cases and the \acronymref{OLUT}{OLUTs} are also compared in the first rows of Table~\ref{tab:metrics}, along with other possibilities discussed later.

\begin{figure}[t!]
\centering
\includegraphics[width=0.98\columnwidth]{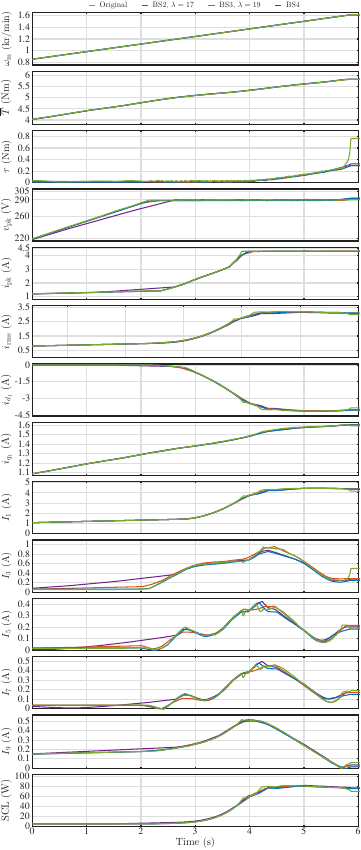}
\caption{Results under a speed ramp from $850$ to $1563\rpm$, for the \acronymref{OLUT}{OLUTs} (green) \cite{Yepes2026TIE}, for the reduced \acronymref{LUT}{LUTs} using \approachref{BS2} with $\lambda=17$ (blue), \approachref{BS3} with $\lambda=19$ (red), and \approachref{BS4} with target \acronymref{LUT}{LUT} size of $1225$ (purple).}
\label{fig:simulation}
\end{figure}

In any case, \approachref{BS1} and \approachref{BS2}, which rely on \approachref{LC1}, require generating the \acronymref{OLUT}{OLUTs} beforehand. This is acceptable if the main goal is to save \acronymref{DSP}{DSP} memory, but not if \acronymref{LUT}{LUT} generation time must also be reduced (see Table~\ref{tab:metrics}). For the latter purpose, \approachref{BS3} may be adopted using \approachref{LC2}, slightly reducing the \acronymref{LUT}{LUT} sizes to $23\times47=1081$ entries and greatly shortening the generation time to $9\min$ (see Table~\ref{tab:metrics}). However, the feasible $T$-$\omega$ range then decreases from the \approachref{BS2} values of $99.4\%$, $100\%$, and $100\%$ for $\epsilon=1\%$, $3\%$, and $5\%$ [diamonds in Fig.~\ref{fig:areaVSsize_zero2extremes}] to $97.2\%$, $98.4\%$, and $99.3\%$, respectively [diamonds in Fig.~\ref{fig:areaVSsize_zero2extremes_zerow2top}]. As shown in Fig.~\ref{fig:PlotArea_zero2extremes_zerow2top_lambda17}, this reduction is mainly concentrated near the positive and negative $T$ limits. For \approachref{BS3}, this can be substantially mitigated using $\lambda=19$, which increases the feasible area to $98.3\%$, $99.9\%$, and $100\%$ [triangles in Fig.~\ref{fig:areaVSsize_zero2extremes_zerow2top}] while further reducing the \acronymref{LUT}{LUT} size to $21\times42=882$ entries. As depicted in Fig.~\ref{fig:PlotArea_zero2extremes_zerow2top}, this improvement over Fig.~\ref{fig:PlotArea_zero2extremes_zerow2top_lambda17} is mainly achieved because the outermost $T$ breakpoints ($\pm19.0$ instead of $\pm18.7\Nm$) are closer to the \acronymref{OLUT}{OLUT} $T$ limits ($\pm19.3\Nm$). Nevertheless, even for a case like that in Fig.~\ref{fig:PlotArea_zero2extremes_zerow2top_lambda17}, it is possible to add feasible $T$ and $\omega$ breakpoints matching the outermost ones of the \acronymref{OLUT}{OLUTs} by using \approachreftext{BS3}{BS3B} without much longer overall \acronymref{LUT}{LUT} generation time. This gives a feasible area [see Fig.~\ref{fig:PlotArea_3B} and Table~\ref{tab:metrics}] similar to \approachref{BS2} ($99.3\%$ versus $99.4\%$) with just different $\omega$ breakpoints [cf. Fig.~\ref{fig:PlotArea_zero2extremes}], in a relatively short time.

\begin{figure}[t!]
\centering
\includegraphics[width=\columnwidth]{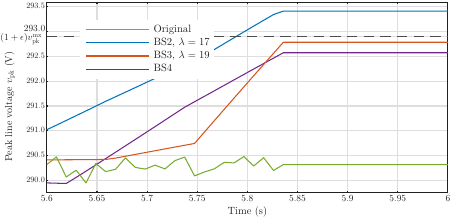}
\caption{Zoom of the final interval of $v_{\mathrm{pk}}$ in the simulation shown in the fourth subplot of Fig.~\ref{fig:simulation}. The dashed line indicates $(1+\epsilon)v^{\mathrm{mx}}_{{\mathrm{pk}}}$ for $\epsilon=1\%$.}
\label{fig:simulation_vpk_zoom_5p6_to_6p0}
\end{figure}

For \approachref{BS1}--\approachref{BS3} and \approachreftext{BS3}{BS3B}, the reduced \acronymref{LUT}{LUT} breakpoints were mostly uniformly spaced, for simplicity. To further mitigate the impact of the \acronymref{LUT}{LUT} reduction on $T$-$\omega$ range, \approachref{BS4} may be used. However, \approachref{BS4}  requires the \acronymref{OLUT}{OLUT} data. As shown at the bottom of Table~\ref{tab:metrics}, e.g., when setting the target \acronymref{LUT}{LUT} size to $1225$ entries and $\epsilon=1\%$, applying the simplification procedure in Fig.~\ref{fig:flowchart_LUT_reduction} takes $975\min$; this is added to the $574\min$ for the \acronymref{OLUT}{OLUTs}, giving $1549\min$ (see Table~\ref{tab:metrics}). This results in \acronymref{LUT}{LUTs} with $39\times31=1209$ entries, whose reduced breakpoint sets are displayed in Table~\ref{tab:nonuniform_breakpoints}.\footnote{BS4 retains rectangular storage but its nonuniform BS needs (at most) $11$ binary-search comparisons ($6$ for $T$, $5$ for $\omega$) per update; the resulting indices and weights serve all $10$ LUTs, while its $70$ $16$-bit breakpoints occupy $140\B$.} Compared with other \acronymref{LUT}{LUTs} of similar sizes  (see Table~\ref{tab:metrics}), the feasible $T$-$\omega$ range is increased to $99.8\%$ [see Figs.~\ref{fig:PlotArea_1209} and~\ref{fig:areaVSsize_approach4}], despite its comparable memory. Moreover, the curve of feasible area versus size is much smoother in Fig.~\ref{fig:areaVSsize} using \approachref{BS4} than other approaches.

Table~\ref{tab:interpolation_method} compares the feasible $T$-$\omega$ areas for the different interpolation methods \cite{MathWorks_griddedInterpolant}, for \approachref{BS2} using $\lambda=17$ (as an example) and $\epsilon=1\%$. Besides being simpler, linear interpolation is clearly the best option regarding feasible area. This was also verified for the other cases in Figs.~\ref{fig:plotting_area} and~\ref{fig:areaVSsize}. This is because the \acronymref{LUT}{LUT} surfaces, as shown in~\cite{Yepes2026TIE}, are largely piecewise smooth, with extended flat regions and sharp slope changes near the activation of inequality constraints. Hence, linear interpolation preserves these features more faithfully, whereas higher-order methods tend to smooth or overshoot them.

To sum up, Table~\ref{tab:metrics} shows that the selected \acronymref{LUT}{LUTs} reduce the total memory of the ten \acronymref{LUT}{LUTs} from $6.1\MB$ to just $18$--$25\kB$ (i.e., by $340$--$250$ times), for any of the \acronymref{BS}{BS} and \acronymref{LC}{LC} approaches. Thus, the storage requirements of all the selected configurations can be expected to be compatible with industrial DSPs. If only the LUT entries of feasible points were stored, the size would be decreased from $4.4\MB$ of the OLUTs to values even lower than $18$--$25\kB$ for the reduced LUTs, albeit at the cost of more complex indexing. On the other hand, the memory savings of the reduced LUTs is achieved while keeping nearly unaltered feasible $T$-$\omega$ area ($97.2$--$99.8\%$), with \approachreftext{BS4}{BS4} particularly excelling in this regard. Concerning offline \acronymref{LUT}{LUT} generation time, it is shortened from $574\min$ to only $9$--$10\min$, i.e., by about $64$ times,\footnote{The original required time of $574\min$ was also given in Supplementary Appendix~B of \cite{Yepes2026TIE}, for the assumed time resolution given by $n_{\mathrm{t}}=150$.} by using \approachref{BS3} or~\approachreftext{BS3}{BS3B} (based on \approachreftext{LC2}{LC2}). In all cases, linear interpolation is preferred.

\section{Simulation Results}
\label{sec:sim}

Simulation results of the \acronymref{PMSM}{PMSM} described in Section~\ref{sec:parameters} are performed in MATLAB under a speed ramp from $850$ to $1563\rpm$ for the same $T$-$\omega$ load curve as the experimental one in \cite{Yepes2026TIE}, so that the mean torque $\overline{T}$, the fundamental $q$-axis current $i_{q_{1}}$ and the fundamental current amplitude $I_{1}$ increase with speed. The green plots in Fig.~\ref{fig:simulation} correspond to using the \acronymref{OLUT}{OLUTs} as in \cite{Yepes2026TIE}. The peak line voltage $v_{\mathrm{pk}}$ rises with speed until the base speed of $1100\rpm$, after which $v_{\mathrm{pk}}$ is kept at the specified limit $v^{\mathrm{mx}}_{\mathrm{pk}}=290\V$. This is achieved by increasing the fundamental $d$-axis current $i_{d_{1}}$, as well as the current harmonic amplitudes $I_{3}$, $I_{5}$, $I_{7}$ and $I_{9}$ of the third, fifth, seventh and ninth harmonics, respectively, along with the \acronymref{SCL}{SCL}. When the peak phase current $i_{\mathrm{pk}}$ reaches its limit $i^{\mathrm{mx}}_{\mathrm{pk}}=4.34\A$, the torque ripple $\tau$ begins to rise moderately to further extend the operation range, without exceeding the limit $\tau^{\mathrm{mx}}=2\Nm$. This behavior aligns well with the design of the \acronymref{OLUT}{OLUTs} in \cite{Yepes2026TIE} to extend the feasible $T$-$\omega$ region. 

The blue, red and purple plots in Fig.~\ref{fig:simulation} are obtained for the reduced \acronymref{LUT}{LUTs} using \approachref{BS2} with $\lambda=17$, \approachref{BS3} with $\lambda=19$, and \approachref{BS4} with target \acronymref{LUT}{LUT} size of $1225$ entries (it reaches $1209$), respectively (see Table~\ref{tab:metrics}). The results provided in Fig.~\ref{fig:simulation} for the \acronymref{OLUT}{OLUTs} (green) and reduced \acronymref{LUT}{LUTs} using \approachref{BS2} and \approachref{BS3} (blue and red) overlap almost perfectly for most of the $T$-$\omega$ range, until $1558\rpm$ is surpassed. The current harmonics using \approachref{BS4} (purple) vary to some extent in that interval, but this is not problematic because the \acronymref{SCL}{SCL}, $\overline{T}$ and $\tau$ are as good as for the other approaches, and $i_{\mathrm{pk}}$, $v_{\mathrm{pk}}$ and $\tau$ do not exceed their respective limits. Once the speed rises above $1558\rpm$, the peak line voltage $v_{\mathrm{pk}}$ obtained with the simplified \acronymref{LUT}{LUTs} increases to around $1\%$ of its limit $v^{\mathrm{mx}}_{\mathrm{pk}}=290\V$, i.e., $1.01v^{\mathrm{mx}}_{\mathrm{pk}}=292.9\V$ (see the zoomed view in Fig.~\ref{fig:simulation_vpk_zoom_5p6_to_6p0}). On the other hand, $v_{\mathrm{pk}}$ for the \acronymref{OLUT}{OLUTs} is kept below that $1\%$ threshold even when the maximum speed $1563\rpm$ is reached. In any case, this performance degradation of the reduced \acronymref{LUT}{LUTs} is arguably small (under $2\%$ $v_{\mathrm{pk}}$ excess), despite the significant \acronymref{LUT}{LUT} simplification, and acceptable in many applications. Alternatively, even if such $v_{\mathrm{pk}}$ excess is deemed unacceptable for a certain application, the consequent reduction in maximum feasible speed is very small (about $5\rpm$ in Fig.~\ref{fig:simulation}). Although not shown to avoid visual clutter, similar results have been obtained for the cases of \approachref{BS1} and \approachreftext{BS3}{BS3B} from Table~\ref{tab:metrics}. 

\section{Conclusions}
\label{sec:conclusions}

This paper has addressed the simplification and optimization of the \acronymref{LUT}{LUTs} needed for a recent method of current-reference generation with minimum losses and torque ripple in the full torque-speed range of six-phase \acronymref{PMSM}{PMSMs} with nonsinusoidal \acronymref{backEMF}{back-EMF}. Different approaches to construct the \acronymref{LUT}{LUTs} (i.e., \acronymref{LC}{LC}), to select the breakpoints (i.e., \acronymref{BS}{BS}) and to perform interpolation were evaluated for this purpose. Four \acronymref{BS}{BS} approaches based on uniformly spaced torque/speed breakpoints (\approachreftext{BS1}{BS1}--\approachreftext{BS3}{BS3} and \approachreftext{BS3}{BS3B}) were considered, as well as an alternative iterative approach with nonuniform breakpoint spacing, \approachreftext{BS4}{BS4}, based on removing, one at a time, the torque or speed breakpoint causing the smallest torque-speed-area loss. Several of them (\approachreftext{BS1}{BS1}, \approachreftext{BS2}{BS2} and \approachreftext{BS4}{BS4}) are based on downsampling existing high-resolution \acronymref{OLUT}{OLUTs} (i.e., \approachreftext{LC1}{LC1}), whereas \approachreftext{BS3}{BS3} and \approachreftext{BS3}{BS3B} (relying on \approachreftext{LC2}{LC2}) can be obtained without those \acronymref{OLUT}{OLUTs}.

Most importantly, the study showed that any of the selected approaches is able to drastically decrease the needed memory. Namely, the required memory was reduced from $6.1\,\mathrm{MB}$ (original) to $18$--$25\,\mathrm{kB}$ by any of them. The \acronymref{LUT}{LUT} generation time was shortened from $574\min$ to $9$--$10\min$ by using \approachreftext{BS3}{BS3} or \approachreftext{BS3}{BS3B} (i.e., \approachreftext{LC2}{LC2}). Meanwhile, even with a stringent $1\%$ allowable deviation in mean-torque error and $1\%$ allowable excess over the specified voltage, current, and torque-ripple limits, the feasible torque-speed area with \approachreftext{BS2}{BS2} and \approachreftext{BS3}{BS3B} decreases by just $0.6$--$0.7\%$. \approachreftext{BS4}{BS4} especially excels in this regard, retaining as much as $99.8\%$ of the original feasible torque-speed range, at the cost of longer \acronymref{LUT}{LUT} generation time.

It has also been found that linear interpolation retains a larger feasible range than other interpolation methods.

Simulation results with the parameters of a real  six-phase \acronymref{PMSM}{PMSM} have confirmed that, despite the simplification, the performance degradation is practically negligible.

Future work could focus on using different breakpoints for the \acronymref{LUT}{LUT} of each harmonic component, or directly identifying regions requiring higher breakpoint density in nonuniform grids without first computing the complete OLUT set.


\end{document}